# Anomalous conductance plateau in an asymmetrically biased InAs/In$_{0.52}$Al$_{0.48}$As quantum point contact

P. P. Das[1], K. B. Chetry[2], N. Bhandari[1], J. Wan[1], M. Cahay[1], R. S. Newrock[2], and S. T. Herbert[3]

[1]School of Electronics and Computing Systems, University of Cincinnati, Cincinnati, Ohio 45221, USA

[2]Physics Department, University of Cincinnati, Cincinnati, Ohio 45221, USA

[3]Department of Physics, Xavier University, Cincinnati, Ohio 45207, USA

## Abstract

The appearance and evolution of an anomalous conductance plateau at 0.4 (in units of *2e$^2$/h)* in an In$_{0.52}$Al$_{0.48}$As/InAs quantum point contact (QPC), in the presence of lateral spin-orbit coupling, has been studied at *T*=4.2K as a function of the potential asymmetry between the in-plane gates of the QPC. The anomalous plateau, a signature of spin polarization in the channel, appears only over an intermediate range (around 3 V) of bias asymmetry. It is quite robust, being observed over a maximum range of nearly 1V of the sweep voltage common to the two in-plane gates. Our conductance measurements show evidence of surface roughness scattering from the side walls of the QPC. We show that a strong perpendicular magnetic field leads to magnetic confinement in the channel which reduces the importance of scattering from the side walls and favors the onset of near ballistic transport through the QPC.

For more than a decade, there have been many experimental reports of anomalies in the quantized conductance of quantum point contacts (QPCs). These anomalies appear at non-integer multiples of $G_0 = (2e^2/h)$, and include the observation of anomalous conductance plateaus around $0.5G_0$ and $0.7G_0$.[1-9] There is a growing consensus that these conductance anomalies are indirect evidence for the onset of spin polarization in the narrow portion of the QPC.[10-16] Recently, we used the lateral spin orbit coupling (LSOC) resulting from the lateral in-plane electric field of the confining potential of a QPC with in-plane side gates, to create a strongly spin-polarized current by purely electrical means[16] *in the absence of any applied magnetic field*. We used a non-equilibrium Green's function (NEGF) analysis to model a small QPC[17,18] and found three ingredients that are essential in generating a strong spin polarization: an asymmetric lateral confinement, a LSOC induced by the lateral confining potential of the QPC, and a strong electron-electron (e-e) interaction. It is important to assess the robustness of the anomalous plateau as a function of the bias asymmetry between the two gates, since two QPCs in series, with an asymmetric LSOC, could be used to build a low temperature all-electric spin valve. In the latter, by flipping the polarity of the gates, either a largely spin-up or spin-down electron current can be injected using a single QPC[16,17]. In such a device a large current (due to either spin-up or spin-down electrons) will flow through the spin valve when both QPCs are biased with the same polarity on their two gates (ON condition) and a much smaller current will flow when the polarities are opposite (OFF condition). Towards that goal, we investigate for the first time, the evolution of an anomalous plateau as a function of the bias asymmetry between the two gates of an InAs-based QPC in the presence of LSOC. The anomalous plateau appears only over a limited range of bias asymmetry. It is quite robust, being observed over a maximum sweep voltage range of nearly 1V. (The sweep voltage is common to both of the in-plane gates.)

We provide a possible explanation for the observation of such a conductance anomaly – one that is observable only over a limited range of bias asymmetry.

An InAs/In$_{0.52}$Al$_{0.48}$As modulation-doped symmetric heterostructure, grown by molecular beam epitaxy, was used to fabricate the QPC device. The layer sequence of the heterostructure is shown in Fig.1(a). Shubnikov-de Haas and quantum Hall measurements performed on a simple Hall bar structure yielded the carrier concentration and electron mobility of the two-dimensional electron gas (2DEG), 2.2×10$^{16}$/m$^2$ and 3.67 m$^2$/Vs, respectively. After the sample was cleaned in hot acetone, methanol and isopropanol (for 10 min each), it was washed in an oxygen plasma for 40s. Next, it was pre-etched in 4% HCl for 5 min., flushed in DI water, and pre-baked at 185 ºC for 5 min.[19] A 50 nm thick polymethylcrylate (PMMA) electron beam resist was spin-coated and exposed, using electron beam lithography, to write the QPC pattern. The QPC constriction was defined by etching two trenches about 50 nm deep and 315 nm wide. In the sample used for the results presented here, the narrow portion of the QPC channel has a width and length of 270 nm and 525 nm, respectively. The pattern was then developed in MIBK:isopropanol (1:1) for 65s. After post-baking the sample at 115$^0$C for 5 min., it was etched in H$_2$O:H$_2$O$_2$:CH$_3$COOH (125:20:10) for 25 s. Ohmic contacts were deposited using 12 nm Ni, 20 nm Ge and 300 nm Au, followed by a rapid thermal annealing at 350 $^0$C for 120 s.

The electrostatic width of the QPC channel was changed by applying bias voltages to the metallic in-plane side gates, depleting the channel near the side walls of the QPC. Battery operated DC voltage sources were used to apply constant voltages $V_{G1}$ and $V_{G2}$ to the two gates. An asymmetric potential $\Delta V_G = V_{G1} - V_{G2}$ between the two gates was applied to create spin polarization in the channel[16]. The QPC conductance was then recorded as a function of a common sweep voltage, $V_G$, applied to the two gates in addition to the potentials $V_{G1}$ and $V_{G2}$, with the current (Fig. 1b) flowing in the x-direction. The linear conductance $G$ ($=I/V$) of the channel was measured for different $\Delta V_G$ as a function of $V_G$, using a four-probe lock-in technique with a drive frequency of 17 Hz and a drain-source drive voltage of 100 µV. All measurements were made at $T= 4.2$K. For all values of $\Delta V_G$, the gates were found to be non-leaking.

Figure 2 shows the conductance of the QPC as a function of the sweep voltage $V_G$ for different asymmetric biases ($\Delta V_G=V_{G1}-V_{G2}$) between the gates. In Fig.2, the left-most curve shows the conductance for the symmetric case, i.e., with only the common sweep voltage $V_G$ applied to the gates. For the other curves, from left to right, the potential $V_{G2}$ applied to gate G2 is fixed at -2.0V and the potential $V_{G1}$ on gate G1 is varied from -0.1 to -6.5 V, the latter corresponding to a large asymmetry between the two gates. As can be seen from Fig. 2, an anomalous plateau (around 0.4 $G_0$) is only observed for an intermediate range of bias $\Delta V_G$. Quite remarkably, it appears over a maximum sweep voltage range of nearly 1 V.

The reason for applying negative potentials $V_{G1}$ and $V_{G2}$ on the gates was to push the electrons away from the side walls and reduce the effects of surface roughness scattering. Indeed, a three-dimensional atomic force microscope (AFM) scan of the QPC device, shown in Fig. 1(b), indicates that the side walls of the QPC are rather ragged (especially in the vicinity of the constriction), a result of the wet etching process used to define the trenches. Surface scattering from this roughness is expected to play a critical role in explaining the conductance data, for which we provide the following explanation.

The evolution of the anomalous plateau with $\Delta V_G$ is interpreted as follows. The left-most curve in Fig. 2 corresponds to the symmetric case, i.e., there is only the common sweep voltage $V_G$ applied to the gates. The conductance curve is rather smooth, with no major features at 0.4 or 1.0 $G_0$. We attribute this to significant elastic scattering in the narrow portion of the QPC, due either to surface roughness scattering or dangling bonds at both channel/vacuum interfaces, as supported by the surface ruggedness around the QPC observed using the AFM, Fig.1(b). Since it has been known since the early 1990s that impurity scattering eventually leads to the disappearance of the integer conductance plateaus in symmetrically biased QPCs,[20-25] the absence of clear features in the conductance for a symmetric bias is indirect evidence of surface scattering from the side walls. Hence we apply a negative bias to both in-plane gates when studying the influence of the bias asymmetry. These negative biases push the electrons away from the walls making transport through the channel more ballistic. The asymmetric bias eventually leads to spin polarization in the channel, triggered by the imbalance of the LSOC on the two sides of the channel, as discussed in detail in our earlier work[16]. Despite the different negative biases ($V_{G1}$ and $V_{G2}$) applied to the gates to create the asymmetry, when the sweep voltage increases, the total potential on either gate will eventually approach zero and the importance of surface scattering will reappear. This assertion is supported by the fact that there is still some evidence of elastic scattering in the channel (the normal conductance plateau is not exactly at $G_0$). The fact that the conductance anomaly is located at 0.4 $G_0$ and not 0.5 $G_0$ is in agreement with our recent NEGF simulations of the influence of impurity scattering on the location of the anomalous plateau. There we showed that conductance anomalies can appear at values different from 0.5 $G_0$ if the transport through the QPC is not truly ballistic.[26]

When the bias asymmetry is large (with a more negative bias on gate G1), we expect electrons in the channel to be squeezed towards gate G2, increasing the electron concentration on that side of the channel. As a result there is increased screening of the electron-electron interactions in the channel, which inhibits the onset of spin polarization in the QPC.[16,17] In addition, there is some influence of surface scattering on that side of the QPC as electrons are being squeezed towards gate G2.

To further confirm the presence of surface scattering in our sample, we measured the magnetic field dependence of the conductance as a function of the sweep voltage $V_G$. The magnetic field was perpendicular to the device plane or the 2DEG. The sweep voltage was superimposed on gate biases of $V_{G1}$=-1.0 V and $V_{G2}$ =0 V. Figure 3 shows that under the influence of the magnetic confinement, the 0.4 plateau evolves smoothly towards the normal $G_0$. Furthermore, the latter is sharp and well-pronounced. Magnetic confinement therefore helps to diminish scattering from the side walls. Transport through the channel is then near-ballistic and the normal conductance plateau is well-defined. Although it is not shown here, the magnetic field dependence of the anomalous plateau was similar for those values of the asymmetric gate biases for which the anomalous plateau could be seen. The gradual evolution of the conductance anomaly towards the normal conductance plateau, with increasing magnetic field, is also in agreement with the increase in the electron density in the QPC channel as the magnetic field increases.[16] As a result, the effects of electron-electron interactions are diminished, prohibiting the onset of spin polarization in the channel.

Despite the presence of surface roughness scattering in the QPC, the anomalous plateau (Fig. 2) appears over a fairly wide range (around 3V) of gate bias asymmetry and over a maximum sweep voltage range ($V_G$) of nearly 1 V. Therefore, by flipping the polarity of the gates, a largely spin-up or spin-down electron current can be injected using a single QPC and it can be done over a fairly wide range of asymmetric biases. This is of practical importance since the spin polarization in two QPCs in series could be tuned for either spin up or spin down injection, opening the path for the realization of a low temperature (at least) all-electrical spin valve. The resulting ON/OFF conductance ratio of that spin valve could be further controlled by the addition of in-plane gates acting on the middle channel between the two QPCs. Potentially, this spin valve could work at more elevated temperature by reducing the width of the channel and avoid spin coupling between subbands due to LSOC.

**Acknowledgment**

This work is supported by NSF Awards ECCS 0725404 and 1028483. We thank P. Debray for his assistance throughout the experiments, and S. Datta and J.P. Leburton for helpful comments.

**Figures:**

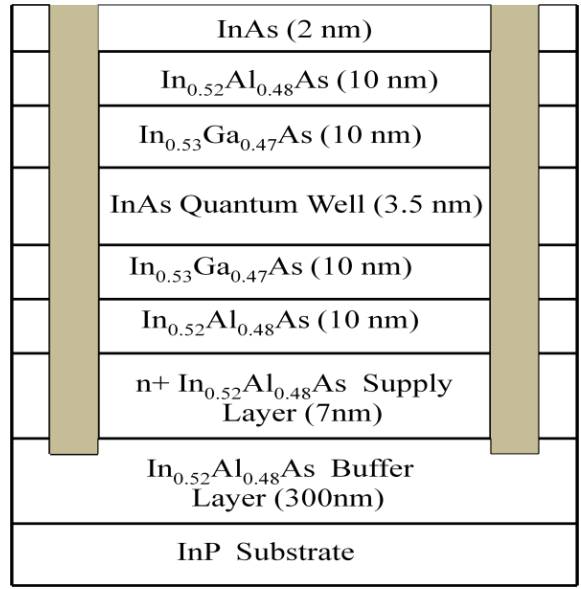

(a)

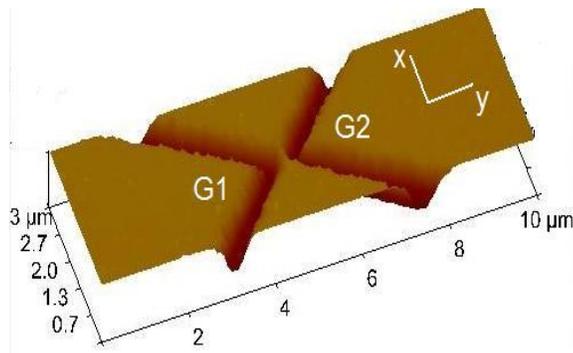

(b)

**Fig. 1** (a). A schematic cross-section of the InAs/In$_{0.52}$Al$_{0.48}$As heterostructure used to make the QPC. The two vertical grey regions represent the trenches used to define the QPC. (b) A three-dimensional AFM image of a QPC with two in-plane gates (G1 and G2), fabricated using a chemical wet etching technique. The current flows in the x-direction. An asymmetric LSOC is generated using an asymmetric bias between the two gates generating an electric field in the y-direction.

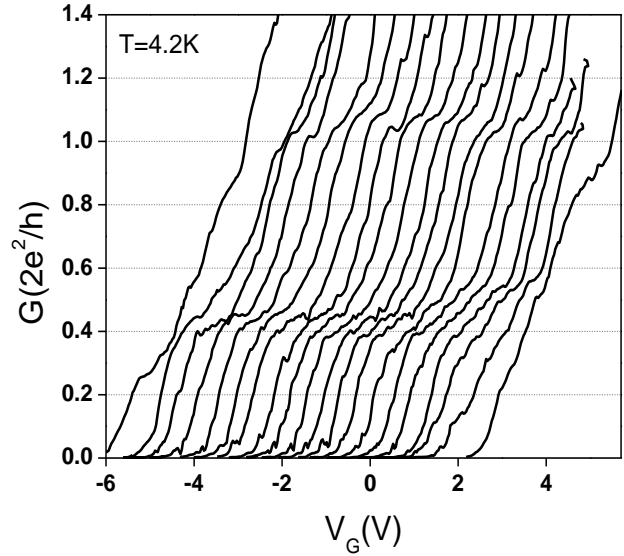

**Fig. 2.** The conductance of the QPC (in units of $2e^2/h$) measured as a function of the commom sweep voltage $V_G$ applied to the in-plane gates, at $T$=4.2K. The sweep voltage is superimposed on initial potentials $V_{G1}$ and $V_{G2}$ applied to the gates to create an asymmetry. The left-most curve shows the conductance for the symmetric case; i.e., with only the common sweep voltage $V_G$ applied to the gates. For the other curves, from left to right, the initial potential $V_{G2}$ applied to gate G2 is fixed at -2.0V and the initial potential $V_{G1}$ on gate G1 is equal to 0.0, -0.1, -0.3, -0.6, -0.9, -1.2, -1.5, -1.8, -2.0, -2.3, -2.6, -2.9, -3.1, -3.4, -3.7, -4.0, -4.5, -5.0, -5.5 and -6.5 V, respectively.

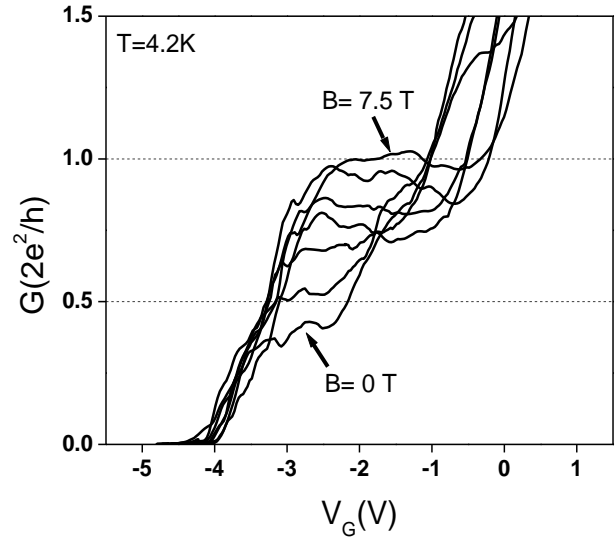

**Fig. 3**. The conductance (in units of $2e^2/h$) in a magnetic field, versus the common sweep voltage $V_G$ applied to the two in-plane gates, at $T$=4.2K. Prior to the application of the common gate bias, the initial biases $V_{G1}$ and $V_{G2}$ were set equal to -1V and 0V, respectively. The different curves correspond to different values of the constant magnetic field applied perpendicular to the 2DEG. The plots corresponding to B = 0T and 7.5T are shown by arrows. From bottom to top, the other curves correspond to magnetic field values of 3.0 T, 5.5 T, 6.0 T, 6.5 T and 7.0 T, respectively.